\documentclass[prl,floatfix,twocolumn,showpacs,fleqn]{revtex4-1}
\usepackage[T1]{fontenc}
\usepackage[english,francais]{babel}
\usepackage{amsfonts}
\usepackage{amsmath}
\usepackage{amssymb}
\usepackage[dvips,pdftex]{graphicx}
\usepackage[latin1]{inputenc}
\usepackage[margin=1 in]{geometry}
\usepackage[dvips,pdftex]{graphicx}
\usepackage{lipsum}

\begin{document}

\title{An interplay of migratory and division forces as a generic mechanism for stem cell patterns}

\author{Edouard Hannezo$^{1,2}$}
\author{Alice Coucke$^{1,3,4}$}
\author{Jean-Fran\c{c}ois Joanny$^{1,5}$}

\affiliation{$^{1}$Physicochimie Curie (Institut Curie / CNRS-UMR168 /UPMC), Institut Curie, Centre de
Recherche, 26 rue d'Ulm 75248 Paris Cedex 05 France, PSL Research University}
\affiliation{$^2$Cavendish Laboratory, 19 JJ Thompson Avenue, CB3 0HE, Cambridge, United Kingdom}
\affiliation{$^3$Laboratoire de Physique Th\'{e}orique, CNRS-UMR8549, Ecole Normale Sup\'{e}rieure, 24 Rue Lhomond, 75005 Paris, France}
\affiliation{$^4$Biologie Computationnelle et Quantitative, CNRS-UMR7238, Sorbonne Universit\'{e}s, UPMC Paris 06, France}
\affiliation{$^5$ESPCI Paris-Tech, 10 rue Vauquelin, 75005, Paris, France}

\begin{abstract}
In many adult tissues, stem cells and differentiated cells are not homogeneously 
distributed: stem cells are arranged in periodic "niches", and differentiated 
cells are constantly produced and migrate out of these niches. In this article, we 
provide a general theoretical framework to study mixtures of dividing and actively 
migrating particles, which we apply to biological tissues. 
We show in particular that the interplay between the stresses arising from 
active cell migration and stem cell division give rise to robust stem cell 
patterns. The instability of the tissue leads to spatial patterns which are either steady or oscillating in time. 
The wavelength of the instability has an order of magnitude consistent 
with the biological observations. We also discuss the implications 
of these results for future \textit{in vitro} and \textit{in vivo} experiments.
\end{abstract}
\maketitle

A fascinating property of developing biological tissues is the ability of 
initially identical cells to differentiate and form robust macroscopic 
patterns. This raises the important question of the transmission of biological  
information on scales much larger than the cell size, and of the origin of the 
positional information. The pioneering work of Turing \cite{1} has shown that 
diffusion-reaction mechanisms are a generic way of to create self-organized 
patterns \cite{ 2, 2b, 5,5b, 7c}, and the discovery of morphogens has stimulated a 
renewed interest for these ideas \cite{7d}. Nevertheless, their applicability in biology 
remains limited, and most of the patterns that have been studied have been shown 
to result rather from short range cellular inhibition than from a long-range 
diffusion gradient \cite{3}. 

On the other hand, mechanical stress is increasingly being recognized as an 
important regulator of tissue homeostasis and of many cellular events such as cell 
division, differentiation and death \cite{8,9,10,11,12,12a}. It is therefore of 
major importance to understand the coupling between the mechanical state of a 
tissue and the regulation and patterning of stem cells. Although long-ranged morphogens gradients play a crucial role in differentiation \cite{7d}, mechanical signaling is 
long-range and propagates fast, which makes it a robust candidate for patterning as well. 
On the other hand,  collective cell migration in cultured cells creates large 
stresses, which propagate on macroscopic scales \cite{13}, in a similar manner to the flocking studied by Toner and Tu \cite{13aa}. Nevertheless, the 
contribution of active cell migration to the homeostasis and patterning of stem 
cells in tissues has not, to the best of our knowledge, been investigated in the 
past. In many instances such as in the skin and 
intestinal epithelia, stem cells are not homogeneously distributed, but are rather 
located in periodically spaced niches, where they divide and give rise to 
differentiated cells. 
Here, we build up a theoretical framework to investigate simultaneously cell 
division, differentiation and active migration in a tissue.  We show that the 
interplay between the stresses generated by division and migration is a generic 
route toward tissue patterning. Indeed, as the differentiated cells actively migrate out of the 
niches, they create stress gradients that maintain the niche under tension, a mechanical feedback that has been shown to enhance cell division and influence stem cell fate. This creates a self-reinforcing loop, where stem cells produce differentiated cells, which migrate away, enabling yet more stem cell divisions. 
\begin{figure*}
	\includegraphics[width=15cm]{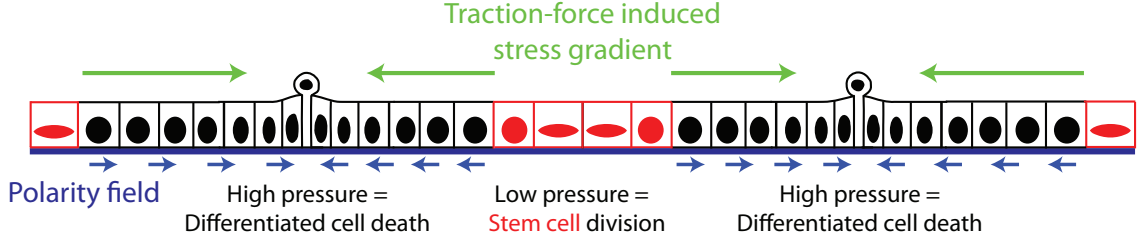}
	\caption{
	Sketch of our model. Differentiate cells actively migrate out of the stem cell compartment, putting it under tension. 
	Stem cell give rise to differentiate cells which actively migrate out of the stem cell compartment. By doing so, they put under tension the stem cell compartment, enabling more stem cell division, and actively maintaining the phase separation.
	}
	\label{schema}
\end{figure*}
We consider only two cell types: stem cells divide and differentiate, but cannot 
actively migrate, whereas 
differentiated cells actively migrate but cannot divide \cite{15}. Cells actively 
migrate by exerting lamellipodial forces on the 
substrate in a polarized manner. We model this migratory polarization by a vector $\bf {p}$, 
pointing towards the front of the cell (i.e. from the center of mass of the cell towards the  lamellipodia). There are therefore four hydrodynamic 
variables: the 
densities of stem and differentiated cells $\rho_s$ and $\rho_d$, the hydrodynamic 
velocity of the mixture $\bf{v}$ and the polarization field $\bf{p}$.
The conservation equations for cell densities read:
\begin{eqnarray}
\partial_t \rho_s +  \nabla (\rho_s {\bf v_s})&=& k(\rho_s, \rho_d) \rho_s - k_d \rho_s \nonumber\\
\partial_t \rho_d +  \nabla (\rho_d {\bf v_d})&=& k_d \rho_s - k_a \rho_d
\label{homeo}
\end{eqnarray}

where $k$, $k_d$ are respectively the division and differentiation rate of stem cells and $k_a$ the loss rate of differentiated cells. At homeostasis, in the absence of cell flow ($v=0$), the cell densities $\rho_s^h$ 
and $\rho_d^h$ are uniform. However, Eqns. \ref{homeo} are unstable if there is no  
negative feedback preventing infinite growth of tissues as soon as 
$k>k_d$. Following previous works, we assume that cell division is 
regulated by the total cell density and expand the division rate at linear order 
around the homeostatic state: 
$$k - k_d = - \frac{1}{\tau_s} (\rho_s + \rho_d - 1)$$
setting $\rho_s^h + \rho_d^h=1$ without loss of generality. Finally, one needs to specify a constitutive equation for the relative flux $J$. We assume it is a diffusive current caused by gradients in the fraction of stem cells: $J =
D \nabla \frac{\rho_s}{\rho_s+\rho_d}$, $D$ being a diffusion constant. In the following, we assume $D = 0$ for the sake of simplicity, although a finite value of D simply shifts the instability threshold without changing the physics of the instability.

We use the theoretical framework of Ref.\cite{16} for 
the dynamics of two cell populations of respective densities and local velocities 
$\rho_\alpha$ and $v_\alpha$ ($\alpha=s,d$).
The barycentric velocity ${\bf v}$ and the relative flux ${\bf J}$ are defined by 
${\bf v_s} = {\bf v} + {\bf J}/\rho_s$ and  ${\bf v_d} = {\bf v} - 
{\bf J}/ \rho_d $.
The barycentric velocity is then determined from force balance: 
\begin{equation}
\partial_{i}\sigma_{i j} = \xi(\rho_s,\rho_d) (v_{j} - V(\rho_s, \rho_d) p_{j})
\label{force}
\end{equation}
where $\sigma_{ij}$ is the stress  tensor, $\xi$ is the friction coefficient of 
the 
cell mixture on the solid substrate. The active migration speed $V$ is written as 
$V = V_0 \phi$, to first order, with $V_0$ and $\phi =  \frac{\rho_d}
{\rho_s+\rho_d}$ respectively the migration speed and fraction of differentiated 
cells. This frictional force assumes that the cells are resting on a solid substrate, which is the case for most epithelial cells resting on a stroma, and that only differentiated cells can actively migrate, a hypothesis which has biological grounds, as discussed in Ref. \cite{15}

On timescales larger than the cell turnover time, tissues have been shown to 
behave as viscous fluids \cite{18}, yielding a constitutive equation for the 
stress: $\sigma_{ij} = - \Pi \delta_{ij} +  2\eta_1 v_{ij} + \eta_2 \partial_k v_{k} \delta_{ij}$,
where $\eta_1$ and $\eta_2$ are the shear and bulk viscosities respectively, $v_{ij}= 1/2 (\partial_i v_j + \partial_j v_i -2/3 \partial_k v_{k} \delta_{ij})$ and $\Pi$ the pressure in the tissue. We redefined $2\eta_1+\eta_2 =\eta$ in the following, which is the only relevant quantity in the one-dimensional case that we study in the main text.
In order to calculate the pressure and the polarization 
fields, we follow Ref.\cite{17} and treat the tissue as a quasi-equilibrium 
mixture close to the homeostatic state. We define the relative dimensionless 
concentrations $\delta \rho_\alpha= \frac{\rho_\alpha - \rho_\alpha^h}
{\rho_\alpha^h}$ and expand the effective energy density, keeping all 
quadratic terms allowed by symmetry and of first order close to equilibrium. 
Moreover, we make the crucial assumption that 
differentiated cells have no spontaneous polarity, and expand the 
energy around ${\bf p} = {\bf 0}$:

\begin{multline}
f = \frac{1}{2\chi_s} \delta \rho_s^2+\frac{1}{2\chi_d} \delta \rho_d^2 + \frac{1}{\chi}\delta \rho_s\delta \rho_d + \frac{\nu}{2}p^2 \\ 
+ \frac{K}{2}(\partial_{\alpha} p_{\beta})(\partial_{\alpha} p_{\beta}) + w(\delta 
\rho_s,\delta \rho_d) \partial_{\alpha} p_{\alpha} 
\label{free}
\end{multline}

$\chi_s$, $\chi_d$, $\chi$ are the compressibilities associated with the mixture, 
$\nu$ is a positive constant, $K$ is the 
Frank constant of the polarization field (using the standard one-constant approximation) and $w(\delta \rho_s,\delta \rho_d)$ can 
be 
expanded to first order around the homeostatic state: $w(\delta \rho_s,\delta 
\rho_d)= w_0 + w_s \delta \rho_s+ w_d\delta \rho_d$. This last term exists by symmetry for polar nematics and couples the polarity field to the density. $w_s$ and $w_d$ are coefficients which, because the system is active, can have any signs, and represent the magnitude of the coupling between polarity and the gradients of stem and differentiated cell densities, respectively.

Then the polarization equation reads: 
\begin{equation}
\gamma \partial_t \textbf{p} =  -\frac{\delta F}{\delta \textbf{p}}=  - \nu \textbf{p} + K \Delta \textbf{p} + w_s \nabla \rho_s + w_d \nabla \rho_d 
\label{polarity}
\end{equation}
where $\gamma$ is a rotational viscosity.

We now rewrite the equations for the evolution of the tissue in dimensionless units using $\tau=\gamma /\nu$ as the time unit and $L=(\eta/ \xi)^{1/2}$ as the length unit.
\begin{multline}
\partial_t \rho_s + \nabla( \rho_s \textbf{v} ) = - (\rho_s + 
\rho_d - 1) \rho_s /\tau_s\\
\partial_t \rho_d + \nabla (\rho_d \textbf{v} ) =   \epsilon\left( 
\rho_s - \rho_s^h \rho_d\right /\rho_d^h)/\tau_s \\
\partial_t \textbf{p} = - \textbf{p} + K \Delta \textbf{p} + (w_d/\nu) 
\nabla \rho_d + (w_s/\nu) \nabla \rho_s  \\
\textbf{v}-V_0 \phi \textbf{p} =  \Delta \textbf{v} - \nabla \rho_s/\chi_s -\nabla \rho_d /\chi_d -W\Delta  {\bf p} 
\label{system1}
\end{multline}
where $1/\chi_\alpha+1/\chi =1/{\tilde \chi}_\alpha$, $\epsilon = \tau_s k_d$,
$(K\xi)/(\nu \eta)= {\tilde K}$, $\tau_s/ \tau = {\tilde \tau}_s$, $V_0\tau / L = {\tilde V}_0$, 
$\chi_\alpha L^2 \xi/ \tau = {\tilde \chi}_\alpha$ and $W = (w_s+w_d)/(L^2 v_1 
\xi) $ and we have omitted all ${\tilde \ }$ symbols. \\

In the following, we assume for a sake of simplicity that the compressibility of 
the two cell types are identical, and define $\chi = \chi_s = \chi_d$. We perform 
a linear stability 
analysis of the homeostatic state of Eqns.\ref{system1}, i.e. $v=0, \bf{p}=0$, $\rho_i =$ constant, by looking for the 
evolution of Fourier modes of wave vector $\bf q$. We define a $3$-dimensional
vector ${\bf X} (\delta \rho_s, \delta \rho_d, {\bf p})$ and denote by 
${\tilde{\bf X}}({\bf q})$ its Fourier transform. The linearization of 
Eqns.\ref{system1} is written in the form ${\partial_t}{\tilde {\bf X}}= M 
{\tilde{ \bf 
X}}$ with M defined as
\begin{widetext}
\begin{equation}
 M = \left( {\begin{array}{ccc}
 - (1-\phi) ( \frac{1}{\tau_s}  + \frac{1}{\chi}\frac{q^2}{1+q^2}) &  - (1-\phi) ( \frac{1}{\tau_s}  + \frac{1}{\chi}\frac{q^2}{1+q^2}) & - \frac{iq}{1+q^2} V_0 \phi (1-\phi) - \frac{i q^3}{1+q^2} W \\
  \epsilon  \frac{1}{\tau_s}  - \phi\frac{1}{\chi}\frac{q^2}{1+q^2} &   - (\frac{1}{\phi}-1)\epsilon \frac{1}{\tau_s} - \phi\frac{1}{\chi}\frac{q^2}{1+q^2}  & - \frac{iq}{1+q^2} V_0 \phi^2 - \frac{i q^3}{1+q^2} W \\
  i q  w_s & i q  w_d & - 1 - K q^2
 \end{array} } \right)
 \label{matrix2}
 \end{equation} 
\end{widetext}

 The signs of the eigenvalues of the stability matrix $M$ determine the 
stability of the tissue.  
If an eigenvalue $s_i (i=1,2,3)$ becomes positive above a threshold $V_c$ of 
migratory speed $V_0$, a perturbation grows at the most unstable wavevector 
$q_c$. 

The solution of this system is rather complicated in the general case, but we 
deduce below the general behavior from special cases and numerical calculations of 
the eigenvalues. The coupling between polarization and density drives the 
instability via two terms : $w_i$ promotes an instability at a finite 
wavelength, and $W$, which is a higher order term, promotes an instability at 
vanishing wavelength ($ q_c\to \infty $) above a threshold $W_c$. This 
instability is unphysical, and higher order terms that would stabilize the tissue 
must be included if $W > W_c$. In the following, we study only the limit $W=0$. 
The results would be qualitatively similar for any $W<W_c$ since, far from the 
threshold, the wavelength of the instability is 
insensitive to the value of $W$.

Moreover, the eigenvalue associated with the polarization is always negative. 
Therefore, we can assume without qualitatively modifying the results, that the 
polarization equation relaxes instantly, and reduce $M$ to  a 
$(2\times2)$ matrix (see SI Text for details). 
Our analysis shows that either one or two eigenvalues of the stability 
matrix can be positive and therefore the bifurcation is of co-dimension 2. We distinguish three types of instabilities:  the homogeneous phase is unstable if 
either only one eigenvalue is real positive, two complex conjugate eigenvalues 
have a positive real part or two eigenvalues are real positive. We plot an illustrative
example of the evolution of the eigenvalues in the Supplementary Text.

Interestingly nevertheless, the wavelength of the stationary spatial instability depends only on the parameters $K$, $\chi$ and $\tau_s$: 
\begin{equation}
\lambda_c = 2 \pi \left(K(\tau_s+ \chi)/ \chi \right)^{1/4}
\label{q}
\end{equation}
In the following, we choose for simplicity the example were $\chi = 1$, $K=1$ and 
$\phi =0.5$. One can give exact expressions for the instability threshold. In order to simplify the analysis, we define $A=\frac{1}{\chi} \frac{q^2}{1+q^2}$ and $B = \frac{V_0 \phi q^2}{(1+K q^2)(1+q^2)}$.

A first transition occurs when only one eigenvalue becomes real 
positive if the active speed exceeds a critical value $V_c^1$, such as $B$ is:
\begin{equation}
V_c^1 \propto B = \frac{1}{ w_d (1+ \epsilon) - w_s (1 - \frac{1-\phi}{\phi} \epsilon) } \left(A+ \frac{1}{\tau_s}\right) \frac{\epsilon}{\phi} 
\label{crit}
\end{equation}

As expected, the instability threshold increases monotonously with the turnover 
rates $\tau_s$. Increasing the coupling $w_d$ decreases the threshold. This criterion also 
indicates that the value $\epsilon_c = \frac{\phi}{1-\phi}$ plays a special role. 
For $\epsilon < \epsilon_c$, increasing $| w_s |$ (if $w_s <0$, i.e. towards more 
negative values) decreases the threshold of the instability. As we expect the 
turnover of differentiated cells to be much faster than the turnover of stem cells 
($\epsilon \ll 1$), this is likely the more realistic limit. In the following, we 
concentrate on this limit $\epsilon < \epsilon_c$. At this transition, the positive 
eigenvalue is real and has no imaginary part, so that we expect formation of 
steady patterns. In the Supplementary Text, we examine the case $\epsilon>\epsilon_c$, which yields to a global phase separation with regions rich in both stem and differentiated cells, and regions poor in both. 
 
A second transition occurs when the two eigenvalues are complex conjugate 
and their reals parts become positive simultaneously. This occurs only above a critical value of the speed $V^2_c$.  At this transition, the presence of two complex 
 conjugate eigenvalues suggest the appearance of spatial patterns oscillating in 
 time, with a frequency given by the imaginary part of the eigenvalue.  For this second, non-stationary bifurcation, the most unstable wavelength $q_2$ is defined by $q_2^4 = \frac{\chi}{K} \frac{1}{\chi+\bar{\tau_s}} $ where $\bar{\tau_s} = \frac{\tau_s}{1-\phi} \frac{\phi}{\phi + \epsilon}$.
One should note the resulting wavelength can become very different from the one of the first transition, in particular if one cell type is in large excess compared to the other ($\phi \to 1$ or $\phi \to 0$).

The critical velocity speed at the threshold $V_c^2$ is such that for $q=q_2$:
$$  V_c^2 \propto B = \frac{1}{w_s \phi + (1-\phi) w_d} \left( A + \frac{1-\phi}{\tau_s} (\frac{\epsilon}{\phi} + 1)  \right)$$

which is always a decreasing function of both couplings $w_s$ and $w_d$.
 
 Importantly, there is always an intersection point between the two transition curves. 
 Indeed, the equation $V^1_c = V^2_c$
always has a solution, which defines a critical point $w_{sc}$ in the $(w_s,V_0)$ space for any value of $w_d$. One can usefully define $\Delta=w_s - w_d$ as the difference of the two couplings, and see that the transition occurs for $\Delta w=0$ (see SI Text).  At this critical point, both eigenvalues have vanishing real and imaginary parts at the same time, characteristic of a Bogdanov-Takens bifurcation.

\begin{figure*}
	\includegraphics[width=10cm]{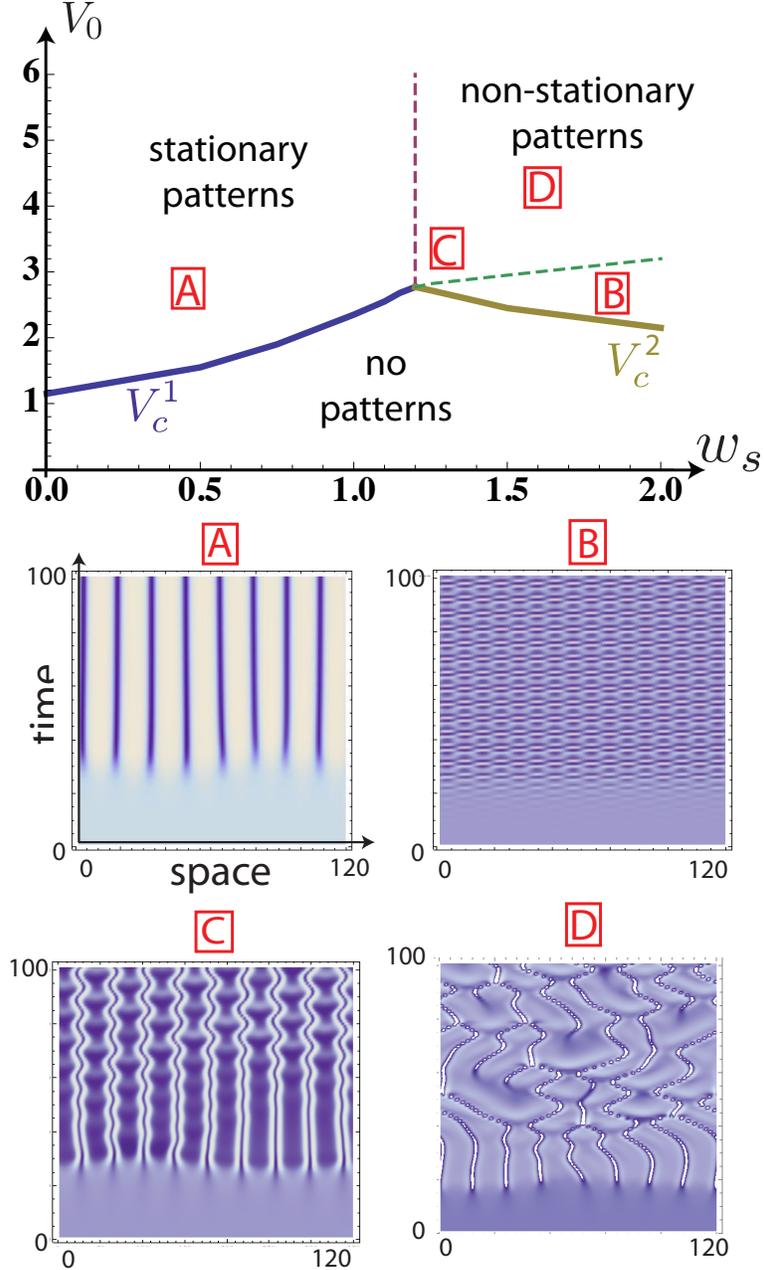}
	\caption{Phase diagram of the patterning instability in 1 dimension. 
	Stationary patterns arises above a critical value of the active migration speed $V_0$, but also non-stationary patterns above a critical value of the coupling $w_s$. We display representative examples of the four phases of the diagram (bellow), showing kymographs of the evolution of the stem cell concentration.}
	\label{num_diagram}
\end{figure*}
 
Moreover, for $\epsilon < \epsilon_c$, $V^1_c$ is an increasing function of $w_s$ whereas 
$V^2_c$ is a decreasing function of $w_s$. Therefore, we expect a reentry 
phenomena due to this non-monotonous threshold for the instability: the 
homogeneous system is stable for intermediate values of $w_s$, and unstable both for small (steady patterns) and large (unsteady patterns) values 
of $w_s$, as can be seen by following a horizontal line in the phase diagram
Fig. \ref{num_diagram}.

We now concentrate on the orders of magnitude of the various parameters. In 
various epithelial tissues \cite{12,13}, the typical migration speed is $V_0 = 10 
\mu m / h$, the typical density is $\rho_0 = 0.01$ cell$/\mu m^2$, the typical 
division time of stem cells is $\tau_s = 24 h$ and the typical fraction of 
differentiated cell is $\phi \approx 0.9$. Moreover, measurements 
suggest $\eta_{3D} \approx 10^5 Pa.s$ \cite{18}  and $\zeta = 10^{10} Pa.m^{-1}$ 
\cite{18a}. If $h=10 \mu m$ is 
the characteristic height of a cell, then $L = (\eta_{3D} h / \zeta)^{1/2} = 
10 \mu m$, agreeing with measurements \cite{19} in the Drosophila scutellum.
Finally, observations on actin dynamics in epithelial sheets and experiments on 
cells 
under tension \cite{16a} suggest a characteristic protrusion time $\tau = 1-2 
min$. The typical compressibility is 
the inverse of the typical pressure exerted by 
tissues: $\chi \approx 10^{-3} Pa^{-1}$ \cite{20}.

We can then evaluate several of the rescaled quantities: $\frac{1}
{\chi} \approx 10$, $\tau_s \approx 10^{3}$, $V_0 \approx 1$. 
Based on available evidence, we suppose 
that the interaction between two polarization vectors is only effective a few cell distances, leading to $K\approx 1$.
This predicts
a critical wavelength of the pattern:
$ \lambda_c \approx 600 \text{ } \mu m$, close to the values observed both in the 
intestinal and skin epithelia \cite{15,2}.
By construction, we 
have the polarity field $\textbf{p}\sim 1$, and since $ \textbf{p} \propto 
w_{s,d} \textbf{q}$, it imposes $w_{s,d} \approx 10$. 
Furthermore, imposing a turnover rate 
$\epsilon \approx 0.1-1$, we can calculate numerically 
the typical value of the critical migration speed, $ V_c \approx 0.1 -1 $,
close to the estimated active migration speed: the stress exerted by differentiated cells is large enough to trigger the instability.

\begin{figure*}
	\includegraphics[width=13cm]{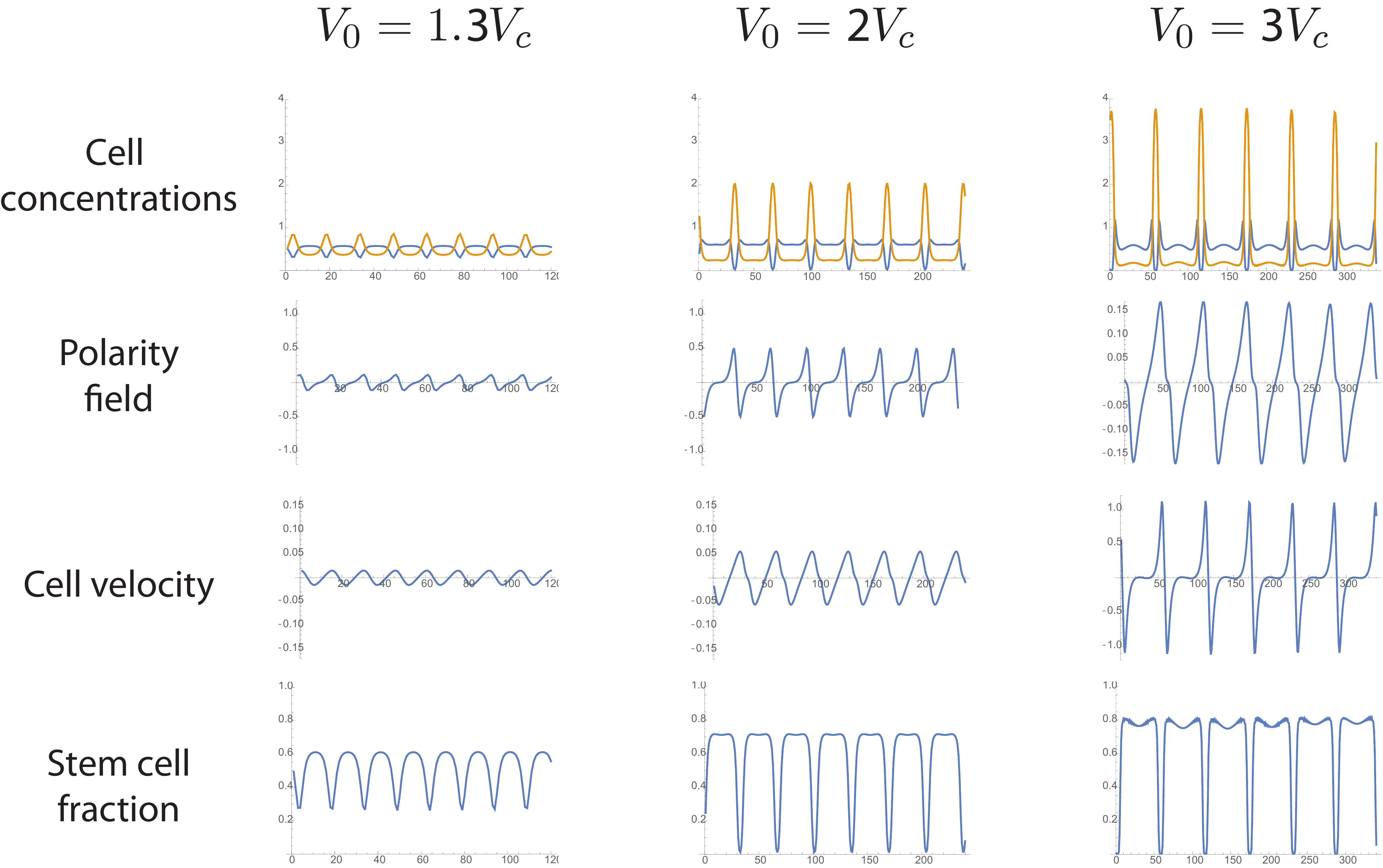}
	\caption{Numerical integration of our equations  in 1 dimension,  for $\epsilon =0.2$, demonstrating the formation of stem cell pattern for high values of the active migration rate $V_0$. We plot the stem/differentiated cell concentration (reps. in orange and blue on the top row) as well as the polarity field $p$, the cell barycentric velocity $v$ and  and the stem cell fraction (in blue on the second, third and fourth row respectively). }
	\label{patterns}
\end{figure*}

\begin{figure*}
	\includegraphics[width=13cm]{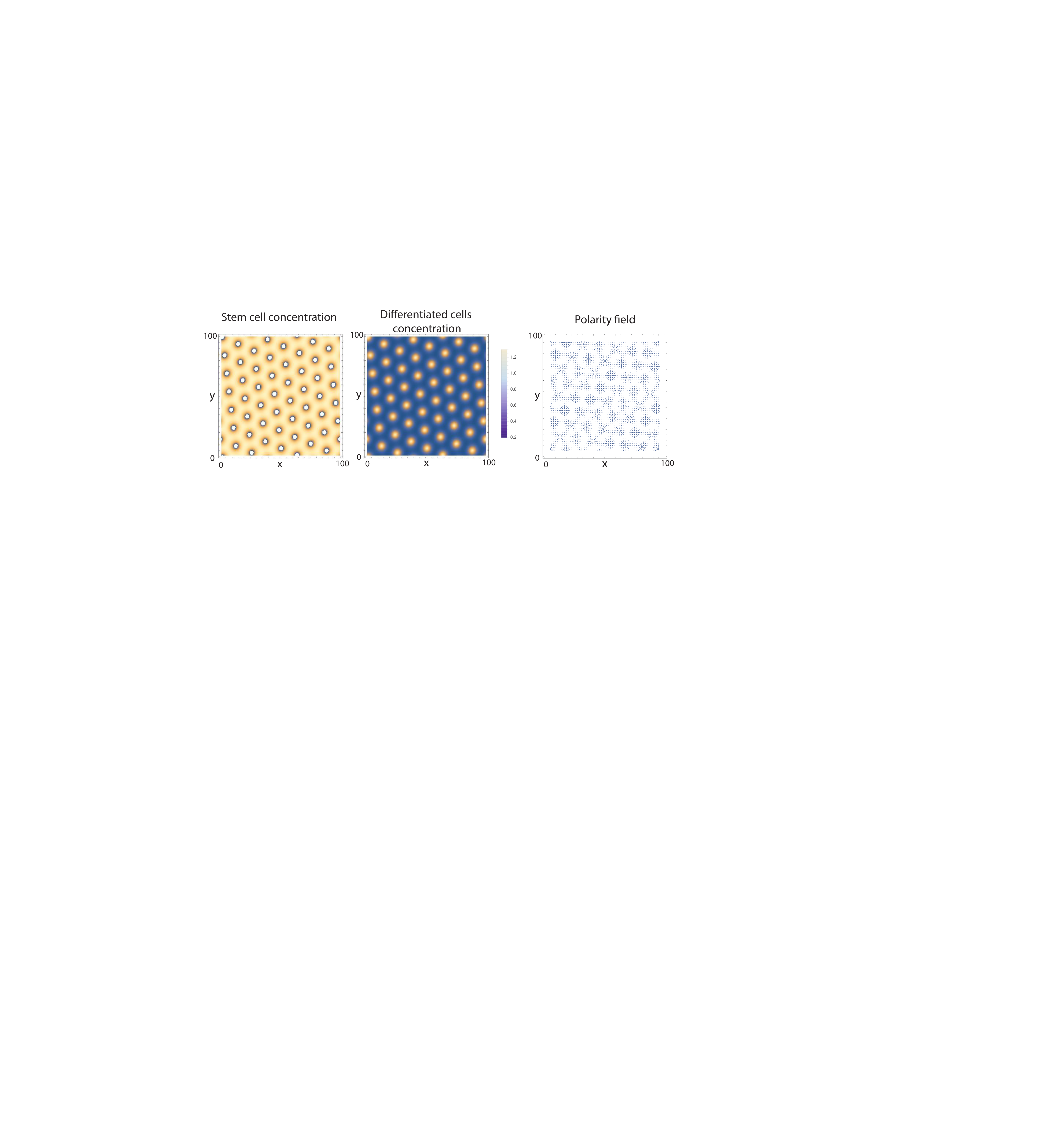}
	\caption{Numerical integration of our equations in 2 dimensions, demonstrating the formation of stem cell pattern for high values of the active migration rate $V_0$. We show a density plot of the stem cell and differentiated cell concentration profile (left and center), and the polarity field (right)  as a function of the two spatial coordinates. The parameters we use, as in Fig. 3 are $\chi = 1$, $K=1$ and $\phi = 1/2$, $\epsilon = 0.2$, $\tau=10$ and $V_0 = 1.2 V_c$.}
	\label{patterns2D}
\end{figure*}

In order to go beyond our stability analysis, we performed a 
full numerical integration of Eq. \ref{system1}. We show the results in 1 dimension  on Fig. \ref{patterns}, as well as in 2 dimensions on Fig. \ref{patterns2D}. We started with $w_s = 0$, $w_d = 1$ and increasing values of the active migration rate $V_0$.  The result is plotted on Fig. \ref{patterns} and displays phase separation and formation of steady state stem cell patterns for high values of $V_0$. The wavelength calculated from this simulation agrees perfectly with the wavelength deduced from our linear stability analysis. The values that we have used for this simulation are $\chi = 1$, $K=1$ and $\phi = 1/2$, $\epsilon = 0.2$, $\tau=10$, with four values of $V_0$ indicated 
on the figure. 
In two-dimensions, hexagonal patterns are observed for all the values we tested, and we show an example for $V_0 = 1.2 V_c$. We neglected the bulk viscosity in the simulation, although we verified that including it did not yield qualitatively different results. We give in Fig. \ref{patterns2D} the intensity plots for the densities of stem and differentiated cells, as well as the polarity field, which goes from stem cell-rich to stem-cell poor regions.

We also explored the 
phase diagram in the plane ($w_s$, $V_0$) for a constant value of 
$w_d = 1.25$, and drew the numerical transition lines from a homogeneous to a patterned state (see  Fig. \ref{num_diagram}), for many values of $w_s$, which agree perfectly with the previous analytical criteria. Patterns can only be seen above a critical value of $V_0$, which depends on $w_s$.
Moreover, below a critical value of the coupling $w_s$, only steady spatial patterns are formed (Fig. \ref{num_diagram}A), whereas above, we observe the appearance of standing waves of finite spatial and temporal frequency (Fig. \ref{num_diagram}B), as predicted by our bifurcation analysis. The numerical transition lines match quantitatively our analytical prediction for $V_c^1$ and  $V_c^2$. 

As shown in Fig. \ref{num_diagram}, there is a complex zoology of non-stationary 
patterns, when the active migration speed is increased further, ranging from caged 
oscillation of stem cell compartments to chaotic motion. These 
patterns are not accessible by a linear stability analysis, and a full non-linear 
treatment would be necessary, but is well beyond the scope of this paper.
Physically, this corresponds to the fact that for large positive values of both 
$w_s$ and $w_d$, the cell polarity field is oriented by two gradients in 
opposite directions ($\rho_s$ and $\rho_d$). Therefore, the cells alternate 
between the two cues, ans this gives rise to complex spatio-temporal patterns (Fig. \ref{num_diagram}C-D).

In this article, we have presented a simple analytical model to describe epithelial tissues and stem cell patterns.  Our model is nevertheless general for any mixture of particles actively dividing and migration, since we considered all hydrodynamical coupling allowed by symmetry, our only assumption being that there was no spontaneous polarisation in the homogeneous state. The patterns we studied reflect a compromise between stresses exerted by the migration of differentiated cells and the division of stem cells. Above a threshold value for the active migration velocity, a positive feedback loop drives the partial phase separation of  the tissue into stem cell-rich and stem cell poor regions, which are either stationary or dynamic in time. This is due to an effective collective migration effect, where migration polarity is coupled to the gradient of cell concentrations in the tissue. Complete phase-separation has been found to be quite general in active self-propelled one-component systems \cite{45}, although here, turnover prevents complete phase separation, yielding robust patterns.

This mechanism is in contrast with previous mechanism of patterning, which rely on diffusion, either of a contractile specie in active fluids \cite{12} or of a morphogen in the classical Turing framework \cite{1}. One advantage of our mechanism is that it does not require a given genetic pathway, but only rather two ingredients that are already known to exist generically in tissues : coupling of neigbouring cells polarity fields, and cell turnover. This could be a source of robustness, in addition to not requiring diffusion of a molecule over long ranges, which might be challenging be achieve in many situations.

In our framework, the pressure in the epithelium would follow the same pattern as the cell concentration. A straightforward extension of our model would therefore be to consider a two dimensional description of the epithelium flows on an arbitrarily curved substrate, and to study the corresponding buckling instability \cite{22b}. This is particularly topical given a recent combination of theory and experiment suggesting that stem cell fate was linked to the local curvature of the epithelium and underlying stroma \cite{22c}.

Our modelling therefore suggests two future research directions. On the one hand, experiments would 
be needed to verify our analytical prediction on the influence of active migration velocity on patterning. In vivo, this could be achieved by inhibiting actin cell migration through specific Arp2/3 inhibitors \cite{22a} to test whether this disrupts stem cell niches. In vivo, traction force microscopy \cite{13} could be used on cultured reconstituted epidermis, which exhibit stem cell patterns \cite{2}. The measured active migration field could then be correlated in time and space with live-markers for stem cells, in order to test our predictions, as well as measuring the coupling constants $w_i$. The existence of oscillating patterns could also be tested in the same system, as well as the dependency of the pattern wavelength on division rate. 
On the other hand, more theoretical work would be necessary to include in our description various non-linear terms which could prove especially important for patterning.

\end{document}